\documentstyle[preprint,revtex,eqsecnum]{aps}

\begin{document}
\draft
\begin{title}
\begin{center}
Frequency dependent conductivity of vortex cores\\
in
type II superconductors
\end{center}
\end{title}
\author{Theodore C. Hsu\cite{byline}}
\begin{instit}
\begin{center}
Centre des Recherches sur les Tr\`es Basses Temp\'eratures,\\
BP 166X, 38042 Grenoble, France.
\end{center}
\end{instit}
\begin{abstract}
The recent optical transmission experiment of Karra\"i {\it et al.}
has probed the localized
quasiparticle excitations of vortices in the superconductor
$\rm{YBa_{2}Cu_{3}O_{7}}$.
In this paper we develop a microscopic
description of single vortex dynamics,
based on the Bogoliubov-deGennes equations and self-consistency
through the gap equation, to
determine the response of vortices to a
transverse time dependent electric field.
It is applicable to the low temperature,
clean, extreme type II limit. In
the limit of large planar mass anisotropy it simplifies. Thus it
may be especially relevent to materials such as ${\rm NbSe_2}$ and
high temperature superconductors.
Of particular interest is the response of the vortex at frequencies
near the minigap, $\Delta^{2}/E_{F}$, where $\Delta$ is the
bulk energy gap and $E_{F}$ is the fermi energy.
A dissipative equation of motion for vortex cores
valid at non-zero frequencies is derived. We give a clear
microscopic meaning to the vortex drag parameter.
The expected dipole transition between quasi-particle
states localized at the core is hidden
because of the self-consistent
nature of the vortex potential. Instead the vortex
itself moves and has a resonance at the frequency of the
transition.
We calculate the conductivity of vortices as a function
of frequency.
A analogy is made to the Mattis-Bardeen result for the
electrodynamic response of bulk superconductors that,
unless translation invariance is broken, single particle
properties are `invisible' to a long-wavelength probe.
However we show that upon adding a translation invariance
breaking term the dipole transition re-appears.
This approach may eventually form the basis
of a microscopic theory of vortex pinning at non-zero frequency.
\end{abstract}
\narrowtext

\section{INTRODUCTION}
\label{sec:INTRO}

The condensed state of type II superconductors
can possess a topological defect, a vortex, at
which there
is a zero in the order parameter.
The existence of these defects is made energetically
favourable upon the application of a sufficiently
large magnetic field or current.
Associated with these defects are quasiparticle
states with wavefunctions localized near the vortex cores
and energies within the bulk energy gap.
These states are superpositions of electrons and holes
and one may think of them as trapped in the vortex
by continual Andreev reflections due to
the spatially varying order parameter.
Their quantitative description in a simple limit
was formulated some time ago.
Caroli, deGennes and Matricon \cite{CAROLIA}
and later Bardeen {\it et al.} \cite{BARDEEN}
calculated the energies and
wavefunctions of these discrete levels
using the Bogoliubov-deGennes (BdG) equations.

The electrodynamical response of type II
superconductors in the mixed state is determined
by the dynamics of vortices. It is usual to
approach this by saying that the phase slip from vortex
motion is associated with voltage drops and hence
gives rise to
dissipation in the superconducting state. Another,
less conventional, way to look at things is to note that
the low energy excitations of the system are associated
with the states inside the bulk energy gap which are
localized near vortex cores. For most practical
applications, however,
vortex pinning is the most important effect.
But in the case of very clean
systems (possibly probed at high frequencies)
one is interested in dissipative flux flow.
The microscopic calculations mentioned above helped form a basis
for theories of vortex motion
based on the idea of a `normal' core such as
those of Bardeen and Stephen \cite{BARDEEN_STEPHEN},
Nozieres and Vinen \cite{NOZIERES} and others \cite{OTHERS}.

Even in the presence of weak pinning, flux flow may be relevant
if pinning is overcome
at high enough temperatures (but below $T_{c}$) or
high enough frequencies.
Gittleman and Rosenblum \cite{GITTLEMANA,GITTLEMANB}
studied dissipation in the mixed state of
conventional superconductors
as a function of frequency and noticed that there
was a crossover to essentially free vortex behaviour
over one or two orders of magnitude centered about some frequency.
What is significant for
this work is that the crossover frequency region appears
to be of the same order of magnitude as the expected
spacing between discrete levels in vortex cores. That spacing
is of order $\Delta^{2}/E_{F}$ or $\hbar^{2}/m\xi^{2}$
(where $\Delta$ is the bulk gap, $E_{F}$ is the Fermi energy
and $\xi$ is the coherence length) and follows from
applying the uncertainty principle to a quasiparticle confined
to a vortex core of size $\xi$. Unless impurities radically
alter the vortex size this should be the characteristic
energy scale. The comparison
is made in table \ref{table} where some unpublished data of
Schleger and Hardy \cite{SCHLEGER} are included. The fact that
the de-pinning and level spacing frequencies are roughly comparable
means that any microscopic understanding of the de-pinning transition
as a function of frequency should take into account the discrete
spectrum of states inside vortex cores.

Caroli and Matricon \cite{CAROLIB} discussed
possible ways of observing the discrete structure
within vortex cores through ultrasonic and
nuclear magnetic relaxation. None of these methods
has so far succeeded.
A few years ago Hess {\it et al.} \cite{HESS}
probed the gross structure of vortices in $\rm{NbSe_{2}}$
by scanneling tunneling microscopy (STM). The results
were qualitatively explained by theories \cite{THEORY}
based upon the Eilenberger equation approach of
Kramer and Pesch \cite{KRAMER}. The STM experiments
however had a resolution of about $0.1 {\rm meV}$ whereas
the energy separation of the discrete levels
is expected to be about $10 {\rm mK}$. STM experiments probing
the vortex cores of high temperature superconductors
(where the energy scale is expected to be of order $1 {\rm meV}$
or more) would be extremely desirable but have not yet been
accomplished.
In passing we note that in charge density wave systems
similar microscopic vortex
structure and possible experimental implications
were proposed by Maki and Huang \cite{MAKI}.

Recent experiments carried out by Karra\"i
{\it et al.} \cite {KARRAI} have measured the optical
transmission of the mixed state of ${\rm{YBa_{2}Cu_{3}O_{7}}}$
at low temperature. They observed a knee in the normalized
transmission coefficient in the frequency range
$50-100{\rm cm}^{-1}$ and attributed it to a quasiparticle
resonance. Some features are of interest. First the
resonance frequency is at least 3 or 4 times higher
than expected from a simple microscopic calculation.
Second, the fitted relaxation rate of the resonance is
very large: of the same order of magnitude as the resonant
frequency itself, $\tau^{-1}\sim 50 {\rm cm}^{-1} \sim 10^{12}{\rm s}^{-1}$.
The is roughly the same order of magnitude as the relaxation rate
of ${\rm{YBa_{2}Cu_{3}O_{7}}}$ as measured by microwave surface
impedance \cite{BONN}. The experiment is performed at a temperature
$T = 2{\rm K}$, over an order of magnitude lower than the resonant
frequency which allows us to deal with only the lowest
energy excitations. One would like to know whether the
observed effect is due to the intrinsic response of free vortices,
or whether it is more important to consider defects, with which
the vortex cores would be highly correlated.

In a previous paper \cite{HSU}
the response of quasiparticles in cores of
vortices (in the clean, low temperature limit) to a long wavelength
electromagnetic field was considered. It was
concluded that the motion of the vortex itself had to be
taken into account because it was a self-consistent
potential. In this paper this idea is re-derived in
a different fashion and expanded in order
to be directly relevant to infrared measurements
in high temperature superconductors.
The paper is organized as follows:
In section \ref{sec:FORMALISM} we introduce
the formalism associated with the Bogoliubov
quasiparticles and the BdG equation. In sections
\ref{sec:DISPLACED} and \ref{sec:MOVING} we
describe displaced and moving vortices respectively
using this formalism. In section \ref{sec:RESPONSE}
we derive the equation of motion,
calculate the response of vortices to a long
wavelength electromagnetic wave and discuss its consequences.
We discuss some effects of pinning in section \ref{sec:PINNING}.
It is followed by a discussion and summary in sections
\ref{sec:DISCUSSION} and \ref{sec:CONCLUSION}.

\section{MICROSCOPIC FORMALISM}
\label{sec:FORMALISM}

The full Eliashberg-Gorkov formalism of BCS theory is, of course,
intractable for the problem of a moving vortex. Instead
we consider the simpler BdG equation for s-wave superconductors.
This equation is valid in the case of a short range
instantaneous pairing interaction. It may also be derived
in the quasiclassical limit $k_{F}\xi >> 1$ where $k_{F}$
is the fermi momentum and $\xi$ is the coherence length.
That limit is the leading order result in the
formulation of Eilenberger \cite{EILENBERGER}.
We shall supplement the BdG
equations with a local gap equation
$\Delta \left({\bf r}\right) =
V\langle c_{\uparrow}\left({\bf r}\right)
c_{\downarrow}\left({\bf r}\right)\rangle$
which determines how the vortex itself
moves (self-consistently).
Of course high temperature superconductors stretch
the assumptions made here. The coherence length is
only about 5--10 lattice constants parallel to the Cu-O
planes so that $k_{F}\xi >>1 $ is barely satisfied.
The discreteness of the lattice may also come into play.
 Real superconductors are
of course non-local and high temperature superconductors
may even have d-type symmetry of the order parameter
(for which the microscopic structure of a vortex has not
yet been calculated).
Also, in general, the pairing interaction may be retarded
but since we are concerned with very
low frequencies relative to the gap,
some relevence to real materials may remain.

The eigenfunctions for Bogoliubov quasiparticles
are two component objects which can be thought
of as the electron and hole amplitudes of
the quasiparticle,
\begin{equation}
\psi({\bf r}) = \left(
\begin{array}{c}
u({\bf r})\\
v({\bf r})
\end{array}
\right).
\label{BDG}
\end{equation}
$\psi$ satisfies a Schr\"odinger equation which is just
the BdG equation,
\begin{equation}
i\hbar{{\partial}\over{\partial t}}\psi ({\bf r}) = \sigma^{z}
\left[{1\over 2m}
({\bf p} - \sigma^{z}{e\over c}{\bf A})^{2}
- E_{F}\right]\psi ({\bf r}) +
\left(
\begin{array}{cc}
0&\Delta ({\bf r})\\
\Delta^{*}({\bf r}) &0
\end{array}
\right)\psi ({\bf r}),
\label{BGD}
\end{equation}
where $\sigma^{z}$ is a Pauli matrix, $E_{F}$ is the chemical
potential, and $\Delta = |\Delta({\bf r}-{\bf r_{0}})|
\exp{(-i\theta({\bf r}-{\bf r_{0}}))}$.
$\theta({\bf r}-{\bf r_{0}})$ is the angle about the
center of the vortex ${\bf r_{0}}$
measured from the ${\hat x}$ axis.
By convention the ${\hat z}$ direction
shall be in the direction parallel to the vortex.

Going to second quantized formalism
the conventional definition \cite{BARDEEN}
of the quasiparticle creation operators is,
\begin{equation}
\left(
\begin{array}{c}
{\Gamma}_{\mu\uparrow}^{\dagger}\\
{\Gamma}_{\mu\downarrow}^{\dagger}
\end{array}
\right)
=\int d^{3}{\bf r}
\left(
\begin{array}{cc}
c_{\uparrow}^{\dagger}({\bf r})
& c_{\downarrow}({\bf r})\\
c_{\downarrow}^{\dagger}({\bf r})
&-c_{\uparrow}({\bf r})
\end{array}
\right)\psi_{\mu}({\bf r}).
\label{QUASIPARTICLE}
\end{equation}
where $\mu$ is an index for the low energy solutions
of this equation and $c_{\uparrow}$, $c_{\downarrow}$
denote spin up and
spin down electron operators respectively. It is conventional
to define the creation operators only for positive
energy states (they are a complete set of states).
However it will be convenient for us to use the complete set of
states which includes both positive and negative energy states
but eliminates the spin degeneracy. That is,
we shall use, for $\epsilon_{\mu} > 0$,
\begin{equation}
\gamma_{\mu}^{\dagger} \equiv {\Gamma}_{\mu\uparrow}^{\dagger}\quad ,
\end{equation}
and for $\epsilon_{\mu} < 0$,
\begin{equation}
\gamma_{\mu} \equiv {\Gamma}_{-\mu\downarrow}^{\dagger}\quad ,
\end{equation}
where $-\mu$ refers to the corresponding time reversed positive energy
state. This is possible because if $(u_{\mu}\ v_{\mu})$ is a solution
for energy $\epsilon_{\mu}$ then $(v_{\mu}^{*}\ -u_{\mu}^{*})$ is
a solution for energy $-\epsilon_{\mu}$. Finally, as we shall see below,
we have a ladder of states from negative to positive
energy which are filled for $\epsilon_{\mu} < 0$.
The inverse transformation from electrons to quasiparticles
is then
\begin{eqnarray}
c_{\uparrow}^{\dagger}({\bf r})
	&= \sum_{\mu}\gamma_{\mu}^{\dagger}u_{\mu}^{*}({\bf r})\quad ,
\nonumber\\
c_{\downarrow}^{\dagger}({\bf r})
	&= \sum_{\mu}\gamma_{\mu}^{\dagger}v_{\mu}({\bf r})\quad .
\label{INVERSE}
\end{eqnarray}

We shall be interested in the extreme type II limit with $H << H_{c2}$.
This can certainly be satisfied in the high temperature superconductors.
Vortices are well separated and there is no quasiparticle
tunneling between vortices
(although see reference \cite{CANEL} for a treatment of that situation).
The magnetic field applied to
create the vortices may be ignored in the BdG equation.
Because the magnetic field is spread out over an area
$\lambda^{2}$ it's importance compared to the phase of the
order parameter is reduced by $\xi^{2}/\lambda^{2}$ where
$\xi$ is the coherence length and $\lambda$ is the
penetration depth.

The low energy eigenfunctions for fixed $k_{z}$, $\mu << k_{F\perp}\xi$,
and the radial coordinate $r << \xi$ are
\begin{equation}
\psi_{\mu}({\bf r}) = \left({k_{F}\over{2\pi\xi L_{z}}}\right)^{2}
e^{ik_{z}z}
\left(
\begin{array}{c}
e^{i(\mu - {1\over 2})\phi}J_{\mu - {1\over 2}}(k_{F\perp}r)\\
e^{i(\mu + {1\over 2})\phi}J_{\mu + {1\over 2}}(k_{F\perp}r)
\end{array}
\right)
\label{EIGENFUNCTIONS}
\end{equation}
with angular momentum index $\mu = \pm{1\over 2}, \pm{3\over 2}, ...$
and $k_{F\perp}$ referring to the Fermi momentum
projected onto the $k_{x}$, $k{_y}$ plane.
At distances of order $\xi$ from the vortex center
the wavefunctions begin to decay
exponentially \cite{CAROLIA,CAROLIB}.
The energies as calculated by Kramer and Pesch \cite{KRAMER},
who analytically accounted for some self-consistency effects
due to the gap equation
$\Delta ({\bf r}) = V\langle c_{\uparrow}({\bf r})
c_{\downarrow}({\bf r})\rangle$, are
\begin{equation}
\epsilon_{\mu} =
{
{2\mu \Delta_{0}^{2}}
\over
{k_{F}v_{F}{\rm cos}^{2}\Theta}
}
{\rm ln}({\pi\over 2}\xi_{0}{\rm cos}\Theta/\xi_{1})
,
\quad {\rm cos}\Theta\equiv k_{F\perp}/k_{F}.
\label{EIGENVALUE}
\end{equation}
The zero of energy is the Fermi energy.

The logarithmic factor is a simple rescaling of energies. It was
derived by Kramer and Pesch and arises from self-consistency of
the vortex solution. At low temperature, energy is favoured
over entropy and so the vortex pinches together.
This increases level spacing, pushing down the energies of occupied
states and reducing the occupation of excited states.
This factor will not be important for this paper
but its effects were explored by
Bardeen and Sherman \cite{BARDEEN_SHERMAN} and
Larkin and Ovchinnikov \cite{LARKIN}.

In this paper, in order to simplify matters, we will suppose
that the Fermi surface is
open and nearly cylindrical. That is,
the system is nearly two dimensional and the dispersion in
the (by convention ${\hat{\bf z}}$) direction
perpendicular to the planes is very small, {\it i.e.} the vortex is aligned
perpendicular to the planes.
This is a rather good approximation for materials such as ${\rm NbSe_{2}}$
and the copper-oxide high temperature superconductors. Moreover
there is another factor which helps.
Near two-dimensionality is equivalent to
$k_{F\perp}=k_{F}{\rm cos}\Theta$ varying little with $k_{z}$.
As one can see from
Eq. (\ref{EIGENVALUE}), the dispersion in the $k_{z}$ direction
enters through the ${\rm cos}\Theta$ in the denominator.
Each angular momentum
level broadens into a band. Nevertheless, the one-dimensional density
of states associated with the variation of this cosine
diverges at $\Theta\sim 0$ and this strengthens
the approximation of neglecting the $k_{z}$ dispersion.

\section{DISPLACED VORTEX}
\label{sec:DISPLACED}

The quasiparticle states are a complete set of states so that once
the density matrix in that basis is specified the state of the system
is defined. Conversely it is possible to describe a displaced
vortex in this basis because the order parameter is completely
specified in terms of the quasiparticles through the gap equation.
By using Eqn. (\ref{INVERSE}) and substituting
into the gap equation we have,
\begin{equation}
\Delta({\bf r}) = -V\sum_{\mu ,\nu}
\langle \gamma_{\nu}^{\dagger}\gamma_{\mu}\rangle)
v_{\nu}^{*}({\bf r}) u_{\mu}({\bf r})\quad .
\label{GAP_EQUATION}
\end{equation}
As an aside recall that in the well-known Ginzburg-Landau solution for the
order-parameter profile of a vortex core in an s-wave superconductor we
have $\Delta(r)\sim r$, as $r\rightarrow 0$. From the series representation
of the Bessel functions, $J_{n}(x)\sim x^{|n|}$ for
small x and integer n. From Eqs. (\ref{EIGENFUNCTIONS})
and (\ref{GAP_EQUATION}) we can see that
this property is satisfied
by the vortex on a microscopic level, but it is only the
$\mu = \pm 1/2$ states that contribute to the linear in r
component of the order parameter. At temperatures higher
than $\Delta^{2}/k_{B}E_{F}$ the occupation
of $\mu = \pm 1/2$ states is about the same
and the linear in r component is washed out
because $J_{-n}(x)=(-1)^{n}J_{n}(x)$.

Suppose the vortex is displaced by a small amount
$\delta {\bf r}_{0}$ in a direction
which makes angle $\phi_{0}$ with the x-axis. Let the occupation
be diagonal before displacement, that is,
$\langle \gamma_{\nu}^{\dagger}\gamma_{\mu}\rangle
=
\delta_{\mu\nu}f(\epsilon_{\nu})$, where $f(\epsilon)$ is
the fermi distribution function. The changes $\delta u_{\mu}({\bf r})
= -\delta{\bf r}_{0}\cdot{\bf\nabla}u_{\mu}({\bf r})$
(and similarly for $v_{\mu}({\bf r})$) may be approximated by
assuming that $u$ and $v$ behave roughly like Bessel functions.
This is explained in detail in the appendix.
In the {\it undisplaced}
coordinates, to linear order in $\delta r_{0}$, and
for small ${\bf r}$,

\begin{eqnarray}
\delta\Delta ({\bf r}) \approx
-&V
{
{\delta r_{0} k_{F\perp}}
\over
{2}
}
\sum_{\nu}
\left(f(\epsilon_{\nu}) - f(\epsilon_{\nu + 1})\right)\qquad\qquad&\nonumber\\
\times&\left[e^{i\phi_{0}}
v_{\nu + 1}^{*}({\bf r}) u_{\nu}({\bf r})
+
e^{-i\phi_{0}}
v_{\nu}^{*}({\bf r}) u_{\nu + 1}({\bf r})
\right].&
\label{DELTA}
\end{eqnarray}

It is important to note that small displacements can be described in
terms of changes in the quasiparticle occupation or density matrix.
The above displacement can be represented by the change
\begin{equation}
\delta\langle \gamma_{\nu + 1}^{\dagger}\gamma_{\nu}\rangle
=
{
{\delta r_{0} k_{F\perp}}
\over
{2}
}
[f(\epsilon_{\nu}) - f(\epsilon_{\nu + 1})]
e^{i\phi_{0}},\quad\forall\nu
\label{DISPLACED_MATRIX}
\end{equation}
(and the Hermitian conjugate)
in the quasiparticle density matrix using the undisplaced basis.

\section{MOVING VORTEX}
\label{sec:MOVING}
In this section we consider a vortex moving with a small velocity
and show, using a Galilean transformation, how it can be
approximately described
by quasiparticle excitations in the stationary frame.
Suppose that the vortex moves with velocity ${\bf v}$. Then
using the invariance of the BdG equation the eigenfunctions
transform like
\begin{eqnarray}
&u({\bf r})\rightarrow u({\bf r}-{\bf v}t)e^{im{\bf v}\cdot{\bf r}}\nonumber\\
&v({\bf r})\rightarrow v({\bf r}-{\bf v}t)e^{-im{\bf v}\cdot{\bf r}}\quad .
\end{eqnarray}
The gap transforms like
\begin{equation}
\Delta({\bf r})\rightarrow
\Delta({\bf r}-{\bf v}t)e^{2im{\bf v}\cdot{\bf r}}\quad ,
\end{equation}
while the chemical potential transforms as
\begin{equation}
E_{F}\rightarrow E_{F} + {1\over 2}mv^{2} -
{e\over c}{\bf v}\cdot{\bf A}\quad ,
\end{equation}
assuming the gauge field ${\bf A}$ is constant.
{}From the usual gauge invariance there is an arbitrariness as to
whether to put the extra phase in ${\bf A}$ or $\Delta$. For what follows
we shall take ${\bf A}=0$.

We wish to evaluate a general component
$\langle \gamma_{\nu + 1}^{\dagger}\gamma_{\nu}\rangle$
of the density matrix in the stationary frame.
But the quasiparticle operators (for which the corresponding
density matrix is taken to be diagonal) in the moving frame
look like
\begin{equation}
{\tilde\gamma}_{\nu}^{\dagger} = \int d^{3}{\bf r}
c_{\uparrow}^{\dagger}({\bf r})u_{\nu}({\bf r}-{\bf v}t)
e^{im{\bf v}\cdot{\bf r}}
+
c_{\downarrow}({\bf r})v_{\nu}({\bf r}-{\bf v}t)
e^{-im{\bf v}\cdot{\bf r}}
\label{MOVE_OP_A}
\end{equation}
for $\nu > 0$ and
\begin{equation}
{\tilde\gamma}_{\nu}^{\dagger} = \int d^{3}{\bf r}
-c_{\uparrow}^{\dagger}({\bf r})v_{-\nu}^{*}({\bf r}-{\bf v}t)
e^{im{\bf v}\cdot{\bf r}}
+
c_{\downarrow}({\bf r})u_{-\nu}^{*}({\bf r}-{\bf v}t)
e^{-im{\bf v}\cdot{\bf r}}
\label{MOVE_OP_B}
\end{equation}
for $\nu < 0$.
Let us consider a vortex which is moving but not
displaced. That is, we take $t=0$ in the above expressions.
Also we retain only terms up to first
order in the velocity.
Using the definitions of Eqs. (\ref{MOVE_OP_A}) and
(\ref{MOVE_OP_B}) and expanding the exponential we can rewrite
the anti-commutator as
\begin{equation}
\left\{
\gamma_{\mu},{\tilde\gamma}_{\nu}^{\dagger}
\right\}
\approx \delta_{\mu\nu} +
\int d^{3}{\bf r}
\psi_{\mu}^{\dagger}({\bf r})
\sigma^{z}(im{\bf v}\cdot{\bf r})
\psi_{\nu}({\bf r})\quad .
\label{ANTI}
\end{equation}
Using the results of appendix, we may evaluate the matrix
element in Eq. (\ref{ANTI}) and find
\begin{equation}
\left\{
\gamma_{\mu},{\tilde\gamma}_{\nu}^{\dagger}
\right\}
\approx \delta_{\mu\nu} +
{
{\hbar vk_{F\perp}}
\over
{2i(\epsilon_{\mu}-\epsilon_{\nu})}
}
\left(
e^{i\theta_{0}}\delta_{\mu,\nu -1}
-
e^{-i\theta_{0}}\delta_{\mu,\nu +1}
\right)\quad .
\end{equation}
where $\theta_{0}$ is the angle that ${\bf v}$ makes with ${\hat{\bf x}}$.

Now we are able to calculate the density matrix for a moving
vortex, but in the basis of eigenstates of a stationary vortex.
To linear order in the velocity we have
\begin{equation}
\langle
\gamma_{\nu}^{\dagger}
\gamma_{\nu\pm 1}
\rangle
\approx
\mp [f(\epsilon_{n})-f(\epsilon_{n\pm 1})]
{
{\hbar vk_{F\perp}e^{\mp i\theta_{0}}}
\over
{2i(\epsilon_{\nu\pm 1}-\epsilon_{\nu})}
}\quad .
\label{MOVING_MATRIX}
\end{equation}

Note the similarity in form of this expression with that of
a displaced vortex, Eq. (\ref{DISPLACED_MATRIX}).
Comparing the two we find that
the corresponding displacement is $\delta r = \hbar v/(\Delta^{2}/E_{F})$
at an angle $\pi/2$ relative to the velocity direction.
The naive gap equation is not
satisfied for a moving vortex! The quasiparticles
are displaced relative to the center of the vortex
and have a non-equilibrium distribution as seen
from the lattice frame.

At this point we should note the effect of dispersion in
the $k_{z}$ direction on the energy denominator
in Eq. (\ref{MOVING_MATRIX}). From Eq. (\ref{EIGENVALUE})
we see that it varies as ${\rm cos}^{-2}\Theta$ as we
change $k_{z}$. Therefore the displacement discussed
above is not the same for states of different $k_{z}$.
In fact it will be clear that the different $k_{z}$
components of the vortex will not move together.
We shall ignore this for now and assume a rigid
motion of the vortex but this is one place where
a relaxation of the two-dimensionality assumption will
complicate matters. Vortex bending is another matter
and will be discussed in section \ref{sec:PINNING}
in conjunction with pinning.

\section{RESPONSE TO EXTERNAL FIELD}
\label{sec:RESPONSE}

\subsection{VORTEX EQUATION OF MOTION}

We shall be interested in the linear response of a vortex
to a uniform time-varying supercurrent or electromagnetic wave.
In a previous paper \cite{HSU} an equation of motion was derived
for a vortex in the presence of this perturbation.
The derivation was a bit cumbersome because it relied
on tracking the evolution of the quasiparticle density-matrix
as a function of time for small times and then extrapolating to
all times. In this paper we shall derive the same
result by an entirely different means which takes advantage
of the results derived in the previous two sections.

Consider a long wavelength
electromagnetic
wave incident in the ${\hat z}$ direction, ${\bf A}$,
with polarization making angle $\theta_{0}$
with the ${\hat x}$ axis and perpendicular to the vortex line
itself. The perturbation to the Hamiltonian
is
\begin{equation}
-{e\over{mc}}{\bf A}\cdot{\bf p}
= -{{i\hbar e}\over{mc}}A
\left\{{\rm sin}\left[\theta -(\theta_{0} + {\pi\over 2})\right]
{\partial\over{\partial r}}
+ {{\rm cos}\left[\theta -(\theta_{0} + {\pi\over 2})\right]}
{\partial\over{r\partial \theta}}
\right\}.
\label{PERTURBATION}
\end{equation}

The vortex will move with some
velocity ${\bf v}_{L}$ in the
presence of a background superfluid velocity
${\bf v}_{S}\equiv -e{\bf A}/mc$ resulting from the applied field.
These velocities will be assumed uniform
in the ${\hat{\bf z}}$ direction (along the length of the vortex).
Such an assumption would be valid if the distance that the
wave penetrates the superconductor
(the smaller of the London penetration
depth and the skin depth) were long and the distance
between pins were also long compared to the coherence length.

First let us present a simple example to illustrate
how we shall calculate the motion of the vortex in this paper.
The basic idea is to calculate time derivatives
of the quasiparticle density-matrix
and use the results of sections \ref{sec:DISPLACED} and
\ref{sec:MOVING} to identify the corresponding motions of the
vortex.

{}From the appendix we find that the matrix element of
Eq. (\ref{PERTURBATION}) is approximately
\begin{equation}
\int d^{3}{\bf r} \psi_{\mu\pm 1}^{\dagger}
(-{e\over{mc}}{\bf A}\cdot{\bf p})
\psi_{\mu}
= {
{e\hbar k_{F\perp}A}
\over
{2mc}
}
e^{\mp i(\theta_{0} + {\pi\over 2})}\equiv W_{\mp}.
\end{equation}
If ${\bf A}$ were time dependent and this were the
only perturbation then we would have
\begin{equation}
{d\over{dt}}
\langle\gamma_{\mu}^{\dagger}\gamma_{\mu -1}\rangle(t)
=
\left[
f(\epsilon_{\mu})
-
f(\epsilon_{\mu-1})
\right]
\times
\left\{
-i{{W_{-}}\over{\hbar}}
\right\}
\end{equation}
and its Hermitian conjugate.

Compare this result with Eq. (\ref{DISPLACED_MATRIX})
and assume that there is a time dependence in
$\delta {\bf r}_{0}$.
Making the identification
$d(\delta {\bf r}_{0})/dt \equiv {\bf v}_{L}$
we see that ${\bf v}_{L}={\bf v}_{S}$.
This is the well known result, that,
in the absence of dissipation
or other forces the vortex moves with the same velocity
as the background superfluid. This is the expected
result from Galilean invariance or from the fact
that in wave mechanics,
a uniform gauge field, ${\bf A}$, boosts the group velocity
of all waves by a velocity ${\bf v}_{S}=-(e/mc){\bf A}$.

Let us now proceed with the derivation of the full equation
of motion. Suppose that the incident
electromagnetic field
${\bf A}$ is time dependent but still uniform.
It will be convenient to perform a gauge/Galilean transformation
which is a boost by velocity ${\bf v}_{S}$ and
eliminates ${\bf A}$.
Under this transformation,
\begin{eqnarray}
u({\bf r})
\rightarrow&
u({\bf r}) e^{i(e/\hbar c){\bf A}\cdot{\bf r}},\nonumber\\
v({\bf r})
\rightarrow&
v({\bf r}) e^{-i(e/\hbar c){\bf A}\cdot{\bf r}},\nonumber\\
\Delta({\bf r})
\rightarrow&
\Delta({\bf r})
e^{2i(e/\hbar c){\bf A}\cdot{\bf r}}.
\end{eqnarray}
The time dependence of ${\bf A}$ rewritten as
${\bf\cal E}={\dot{\bf A}}/c$ results in a perturbation
$\sigma^{z}e{\bf\cal E}\cdot{\bf r}$.
Now we are left with ${\bf A}=0$ and
a vortex which has velocity ${\bf v}_{L} - {\bf v}_{S}$.
The quasiparticle density matrix has off-diagonal elements
\begin{equation}
\langle\gamma_{\nu}^{\dagger}\gamma_{\nu\pm 1}\rangle
=
\mp
[f(\epsilon_{\nu})
-
f(\epsilon_{\nu\pm 1})]
{
{|{\bf v}_{L} - {\bf v}_{S}|k_{F\perp}e^{\mp i\phi_{0}}}
\over
{2i(\epsilon_{\nu\pm 1}-\epsilon_{\nu})}
}\quad .
\end{equation}
The angle $\phi_{0}$ is the angle of
${\bf v}_{L}-{\bf v}_{S}$
with respect to the ${\hat{\bf x}}$ axis.
Here we remark that the Fermion occupation $f(\epsilon_{\nu})$
does not change to linear order in velocity. Because of the
presence of a gap the energy goes quadratically with the velocity.

Now the time evolution of the off-diagonal elements
of the density matrix coming from the difference
$\epsilon_{\nu} \neq \epsilon_{\nu\pm 1}$
corresponds to the motion of the vortex at velocity
${\bf v}_{L}-{\bf v}_{S}$
as derived in section \ref{sec:MOVING}.
However a moving vortex also has a relative displacement
between the order parameter and the quasiparticles relative to
what the gap equation for a stationary vortex would give.
This displacement is
\begin{equation}
\delta r_{0}^{\prime}=
{
{\hbar|{\bf v}_{L}-{\bf v}_{S}|}
\over
{\epsilon_{\nu +1}-\epsilon_{\nu}}
}
\end{equation}
in a perpendicular direction $\phi_{0}^{\prime} = \phi_{0} - \pi/2$.
Thus the displaced vortex itself will affect the quasiparticle density
matrix through a perturbation
$\delta\Delta = -\delta{\bf r}_{0}^{\prime}
\cdot{\bf\nabla}\Delta$.
Its matrix element is
\begin{equation}
W_{\mu\nu} = \int d^{3}{\bf r} ~\psi_{\mu}^{\dagger}({\bf r})
\left(
\begin{array}{cc}
0 &\delta\Delta\\
\delta\Delta^{*}&0
\end{array}
\right)
\psi_{\nu}({\bf r})
\end{equation}\quad .
In the appendix we show that it has the value
\begin{equation}
W_{\mu\nu} = -{
{\delta r_{0}^{\prime}k_{F\perp}}
\over
{2i}
}
\delta_{\mu ,\nu\mp 1}
(\epsilon_{\nu} - \epsilon_{\mu})
e^{\pm i(\phi_{0}^{\prime} + {\pi\over 2})}\quad .
\end{equation}

The contribution of the relative displacement
$\delta{\bf r}_{0}^{\prime}$
to the evolution of the density matrix is
\begin{equation}
{d\over{dt}}
\langle\gamma_{\mu}^{\dagger}\gamma_{\mu -1}\rangle
=
[f(\epsilon_{\mu})-f(\epsilon_{\mu -1})]
\times
\left\{
-
{{iW_{\mu -1,\mu}}\over{\hbar}}
\right\}
=
{1\over 2}i
|{\bf v}_{L}-{\bf v}_{S}|k_{F\perp}
e^{i\phi_{0}}.
\end{equation}
Comparing with Eq. (\ref{MOVING_MATRIX}) which relates velocity to
off-diagonal elements of the density matrix we may
interpret the results in terms of an acceleration
\begin{equation}
a =
|{\bf v}_{L}-{\bf v}_{S}|
(\epsilon_{\nu +1} - \epsilon_{\nu})/\hbar
\end{equation}
at an angle $\phi_{0} - \pi/2$, or, in other words,
\begin{equation}
{\bf a} =
{{\Delta^{2}}\over{\hbar E_{F}}}
({\bf v}_{L}-{\bf v}_{S})
{\bf\times}
{\hat{\bf z}}\quad .
\end{equation}

As for the electric field we require the matrix
element (using the results from the appendix)
\begin{equation}
\int \psi_{\mu}^{\dagger}
e{\bf\cal E}\cdot{\bf r}\sigma^{z}
\psi_{\nu}
\approx
{
{e{\cal E}k_{F\perp}}
\over
{2m
(\epsilon_{\nu +1} - \epsilon_{\nu})}
}
\left(
e^{i\phi_{0}}\delta_{\mu,\nu -1}
-
e^{-i\phi_{0}}\delta_{\mu,\nu +1}
\right),
\end{equation}
where
$\phi_{0}$ is now the angle of ${\bf\cal E}$.
{}From the electric field then we have a contribution
\begin{equation}
{d\over{dt}}
\langle\gamma_{\mu}^{\dagger}\gamma{\mu -1}\rangle
=
[f(\epsilon_{\mu})-f(\epsilon_{\mu -1})]
\times
{
{e{\cal E}k_{F\perp}}
\over
{2im\hbar
(\epsilon_{\nu} - \epsilon_{\nu -1})}
}
e^{i\phi_{0}}.
\end{equation}
The corresponding acceleration, by comparing with
Eq. (\ref{MOVING_MATRIX}) is
\begin{equation}
{\bf a}=-{{e{\bf\cal E}}\over m}= {\dot{\bf v}}_{S}\quad .
\end{equation}

The last influence on the vortex motion we shall consider
in detail is dissipation. Suppose that we are in the lattice
rest frame and looking at vortex moving with velocity ${\bf v}_{L}$.
We see that there is apparently a non-equilibrium off-diagonal
component whose magnitude is proportional to $v_{L}$ and whose
phase gives the direction of ${\bf v}_{L}$.
Let us assume that, due to scattering from things that break
translation invariance such as impurities or phonons, this
off-diagonal component has a lifetime $\tau$ over which it decays
and re-appears as a thermalized (diagonal),
isotropic contribution to the
density matrix (i.e. not carrying current).
In other words we assume that the core can maintain thermal
equilibrium with the lattice unlike, say, the case of $^3{\rm He}$
in which the cores of vortices can heat up. The parameter $\tau$
which will appear in the vortex equation of motion in the drag term
and in the conductivity has a clear microscopic meaning. It is the
lifetime of low energy quasiparticle states.

This theory of the dissipation then gives an additional component
to the time derivative of the vortex velocity,
\begin{equation}
{\dot{\bf v}}_{L} = -{1\over\tau}
{\bf v}_{L}\quad .
\end{equation}
Putting all the contributions together we have the
final equation of motion,
\begin{equation}
\dot{{\bf v}}_{L}
=
\dot{{\bf v}}_{S}
+
{{\Delta_{0}^{2}}\over
{\hbar E_{F}}}
({\bf v}_{L}- {\bf v}_{S}){\bf\times}{\hat z}
-
{1\over\tau}
{\bf v}_{L}.
\label{MOTION}
\end{equation}

It is useful to compare the present expression with that
of Nozieres-Vinen \cite{NOZIERES}. Their equation is derived by
balancing {\it forces}. There is the `magnus' force
$ (hn/2)({\bf v}_{S}-{\bf v}_{L})\times{\hat z}$
where $n$ is the (superfluid) number density of
electrons. Comparing this to the corresponding coefficient
in the present equation of motion, taking
$n=k_{F\perp}^{2}/2\pi$ and
the in-plane coherence length
$\xi_{\perp} = \hbar v_{F\perp}/\pi\Delta$,
one may extract a `mass' of the vortex
$M \sim m_{\perp}(k_{F\perp}\xi_{\perp})^{2}/4$
per unit length in the ${\hat{\bf z}}$ direction.
This expression is perhaps a microscopic justification
for the concept of a `normal core' of size
$\xi$, even at very low temperatures when there
is a gap in the density of states of single particles.
This is should be contrasted with another definition
of the mass of a vortex as recently discussed by Duan and Legget
\cite{DUAN}. This definition is based upon the change in energy
of a moving vortex at order $v^{2}$ which has been
ignored in this paper.

Let us pause and summarize the situation.
We began by asking whether the application of
a long-wavelength electromagnetic field could
cause dipole transitions between the discrete
quasiparticle states in the cores of vortices.
Bearing in mind the approximation of rigid vortex motion
and other assumptions we find that
an applied electric field, instead of causing
dipole transitions, causes the density matrix
to evolve off-diagonal matrix elements
corresponding to vortex motion itself
(after applying the gap equation).
The vortex does not stand still and allow a dipole transition
to take place, as the core of an atom. The difference
is that {\it the vortex
is a self-consistent potential}. Thus the
experimental observation of such quasiparticle
resonances will not be a simple dipole resonance
(except in the presence of pins which will be discussed
in section \ref{sec:PINNING}).
The STM tunneling experiment differs from the present
case in that the tunneling process introduces
an extra particle which is not correlated with the
quasiparticles in the vortex whilst the effect
of an applied electric field is to change the
off-diagonal components of the density matrix.

A simple understanding of the equation of motion
comes from looking at the homogeneous solution.
It can be written
\begin{equation}
\left(
\begin{array}{c}
v_{Lx}\\
v_{Ly}
\end{array}
\right)
=
e^{-t/\tau}
\left[
a_{+}e^{i\Omega_{0}t}
\left(
\begin{array}{c}
1\\
i
\end{array}
\right)
+
a_{-}e^{-i\Omega_{0}t}
\left(
\begin{array}{c}
1\\
-i
\end{array}
\right)
\right]
\label{HOMOGENEOUS}
\end{equation}
which corresponds to moving in circles,
with a definite sense of rotation, at
a frequency $\Omega_{0}\equiv \Delta^{2}/E_{F}$. This is the
sort of state one hopes to excite with
the external probe.
Therefore experimentally it would be
important to probe this system with polarized waves
In the next subsection we discuss what happens when we do so.

\subsection{DISSIPATION IN A SINGLE VORTEX}
\label{subsec:SINGLE}

It is very illuminating to calculate the dissipation in
a single vortex as a function of frequency and polarization
for constant
magnitude $v_{S}$. In the next subsection we shall calculate
the conductivity for a finite density of vortices.
First it is necessary to extract the paramagnetic current
carried by the vortex core. The paramagnetic current is
\begin{equation}
{\bf J}({\bf r})
=
\sum_{\sigma}
c_{\sigma}({\bf r})^{\dagger}
c_{\sigma}({\bf r})\quad .
\end{equation}
Substituting for the electron operators
using the inverse transformation Eq. (\ref{INVERSE}),
and using the results of the appendix to evaluate
matrix elements of $\bf\nabla$ we find that
\begin{equation}
\int {\hat{\bf n}}\cdot{\bf J}
=
{
{ik_{F\perp}}
\over
{2m}
}
\sum_{\nu}
\left(
\gamma_{\nu + 1}^{\dagger}\gamma_{\nu}e^{-i\phi_{0}}
-
\gamma_{\nu}^{\dagger}\gamma_{\nu +1}e^{i\phi_{0}}
\right)\quad ,
\label{CURRENT}
\end{equation}
where ${\hat{\bf n}}$ is some arbitrary unit vector
pointing in direction $\phi_{0}$.

Now we may calculate the (spatially averaged or
zero momentum) paramagnetic current
associated with a vortex motion by comparing Eqs. (\ref{CURRENT})
and (\ref{MOVING_MATRIX}).
It is
\begin{equation}
\langle
\int {{{\bf v}_{L}}\over{v_{L}}}
\cdot
{\bf J}({\bf r})
\rangle
=
{
{v_{L}\hbar^{2}k_{F\perp}^{2}}
\over
{2m(\Delta^{2}/E_{F})}
}
\label{PARAMAGNETIC}
\end{equation}
per unit length in the ${\hat{\bf z}}$ direction
(for a non-cylindrical Fermi surface we would
average over $k_{z}$).

Let us calculate now the dissipation as a function
of frequency for a single vortex.
For simplicity we
shall keep the amplitude $v_{S}$ constant but we shall
allow it to have an arbitrary polarization.
The steady state solution of Eq. (\ref{MOTION}) for
${\bf v}_{S}(t)=
{\bf v}_{S}(0)\exp{(i\omega t)}$
is
\begin{equation}
v_{Ly} - v_{Sy}
=
\left[
{{
\Omega_{0}\tau v_{Sx} - (1 + i\omega\tau)v_{Sy}
}\over{
(1 + i\omega\tau)^{2} + (\Omega_{0}\tau)^{2}
}}
\right],
\label{PARTICULAR}
\end{equation}
and another equation with x,y interchanged and
$\Omega_{0}\rightarrow -\Omega_{0}$.

Eq. (\ref{MOTION}) at zero frequency was introduced by
deGennes and Matricon \cite{DEGENNES}. This had
the drawback of not allowing for small conductivities
seen in flux flow experiments \cite{BORCHERDS} and
was discarded by Nozieres and Vinen in favour of
an equation of motion
where the dissipation acts
on ${\bf v}_{S}$ rather than ${\bf v}_{L}$.
Nevertheless the solution Eq. (\ref{PARTICULAR})
agrees with the Nozieres-Vinen
equation \cite{NOZIERES} at low frequency and
low dissipation. It is probably not valid except in
the clean limit where the levels are well defined.
Moreover, because of the the frequency dependent
term there is a `resonance' at $\omega \approx \Omega_{0}$ with
a width $\tau^{-1}$. If one simply added a ${\dot{\bf v}}_{S}$
term to the Nozieres-Vinen equation one would obtain an
unphysical divergence at that frequency. In addition, even
though Eq. (\ref{MOTION}) was not derived in the
large dissipation limit, that limit
makes sense for the present equation of
motion. The vortex simply stops moving. The Nozieres-Vinen
equation and its relatives have divergent behaviour because
in these models the current through the vortex core is
forced to be equal to the background transport velocity ${\bf v}_{S}$.

There are two contributions to the dissipation which
come from a current (either the background superfluid or
that due to vortex motion) that is in phase with an electric field
(time derivative of the supercurrent or transverse motion
of the vortex).
The first is transverse vortex motion
in phase with the supercurrent.
For clarity let $v_{Sy}=v_{Sx}\exp{i\theta}=v_{S}/\sqrt{2}$.
The parameter $\theta$ simply controls the polarization.
There is an induced voltage per vortex
${\hat z}\times {\vec v}_{L}(h/2e)$. The supercurrent density
$v_{S}ne$ gives dissipation
$-N_{v}(hn/2){\rm Re}(v_{Lx}^{*}v_{Sy} - v_{Ly}^{*}v_{Sx})$,
where $N_{v}$ is the vortex density.
Using Eq. (\ref{PARTICULAR})
\begin{eqnarray}
{\rm Re}(v_{Lx}^{*}v_{Sy} - v_{Ly}^{*}v_{Sx}) &=
-v_{S}^{2}\Omega_{0}\tau
{
{1-(\omega\tau)^{2}+(\Omega_{0}\tau)^{2}}
\over
{[1-(\omega\tau)^{2}+(\Omega_{0}\tau)^{2}]^{2}+4(\omega\tau)^{2}}
}\nonumber\\
&+ {\rm sin}\theta v_{S}^{2}\omega\tau
{
{1+(\omega\tau)^{2}-(\Omega_{0}\tau)^{2}}
\over
{[1-(\omega\tau)^{2}+(\Omega_{0}\tau)^{2}]^{2}+4(\omega\tau)^{2}}
}.
\label{SLIP}
\end{eqnarray}
This first term contributes at low frequencies (relative to
$\Delta^{2}/E_{F}$) and corresponds to the usual phase slip
dissipation mechanism for flux flow.

The second source is the current due
to vortex motion which is in phase and parallel with the applied
electric field. A straightforward calculation gives the average
current density due to vortex motion
to be
$2v_{L}(E_{F}/\Delta)^{2}N_{v}e$ per unit length
in the ${\hat{\bf z}}$ direction.
With the electric field ${\cal E}=(m/e){\dot {\vec v}}_{S}$ the
dissipation is
$2m(E_{F}/\Delta)^{2}N_{v}
{\rm Re}(v_{Ly}^{*}i\omega v_{Sy} +
v_{Lx}^{*}i\omega v_{Sx})$.
Using Eq. (\ref{PARTICULAR}), we obtain \cite{MISTAKE}
\begin{eqnarray}
{\rm Re}(v_{Ly}^{*}i\omega v_{Sy} +
v_{Lx}^{*}i\omega v_{Sx})
=&
v_{S}^{2}
\omega^{2}\tau
{
{1+(\omega\tau)^{2}-(\Omega_{0}\tau)^{2}}
\over
{[1-(\omega\tau)^{2}+(\Omega_{0}\tau)^{2}]^{2}+4(\omega\tau)^{2}}
}\nonumber\\
-&{\rm sin}\theta v_{S}^{2}
\Omega_{0}\omega\tau
{
{1-(\omega\tau)^{2}+(\Omega_{0}\tau)^{2}}
\over
{[1-(\omega\tau)^{2}+(\Omega_{0}\tau)^{2}]^{2}+4(\omega\tau)^{2}}
}.
\label{DRUDE}
\end{eqnarray}

This second part contributes mostly to high frequencies.
It may be understood better by
letting $\Omega_{0}\rightarrow 0$.
In that case we recover a Drude-like expression
$1/[1 + (\omega\tau)^{2}]$ multiplied by the volume of
the vortex cores. This analogous to the well known
zero frequency result for flux flow which is that as
$H\rightarrow H_{c2}$, the resistance in the vortex cores
multiplied by the core volume
matches the resistance for the normal state above $H_{c2}$.

For clarity and in order to add together these two terms
we shall make the oversimplified but well-defined
choice that superfluid density be
equal to the density of all electrons,
$n = k_{F\perp}^2/2\pi$ (again, per unit length in
the ${\hat{\bf z}}$ direction) and
$E_{F}=\hbar^{2}k_{F\perp}^{2}/2m$ independent of $k_{z}$.

With this definite choice we may then plot the dissipation.
In Fig. (\ref{DISSIPATION_PLOT}) we have plotted the dissipation for
right and left circularly polarized ${\bf v}_{S}$. For one of the
polarizations we see dissipation but no special behaviour
near the characteristic frequency of about $\Omega_{0}$
(depending on the magnitude of $\tau$). In the other polarization
the vortex does respond at the characteristic frequency
but essentially as an anti-resonance. There is a {\it minimum}
in dissipation.

\subsection{CONDUCTIVITY}

In this subsection we calculate the experimentally accessible
conductivity at momentum $q=0$. The usual straightforward
way of calculating the conductivity
would be to take the gauge invariant current-current
correlation operator, re-express it in terms of Bogoliubov
operators using Eq. (\ref{INVERSE}) (allowing for time
dependence of the fermion operators) and evaluate its expectation
value. As usual the (time) Fourier transform of this object will
have poles at the excitation frequencies of the system.
The lowest quasi-particle-quasi-hole
excitation would contribute a pole ostensibly around $\Omega_{0}$.
That would lead to a contribution proportional to
$\delta(\omega-\Omega_{0})$
in the real part of the conductivity.

However that is not the whole story because of the `residual'
or `final-state' interaction between these quasiparticles.
It can modify the energy from what one expects given
the effective single-quasiparticle energy spectrum.
Very generally, when symmetry breaking
and long-range order occurs in a Fermionic
system resulting in an energy gap for single-particle
excitations these residual interactions can mix
particle-hole excitations with the collective
(Goldstone) mode
by creating a bound state in the particle-hole channel.
Another way to look at this is to say that while
a mean field or Hartree order parameter is constructed
so that {\it single} particle excitations
have positive energy and do not scatter, when
a particle {\it and} a hole are present the
change in the order parameter due to the hole
affects the particle and {\it vice versa}.

Since the position of a vortex is arbitrary (the real magnetic
field that creates it is arbitrarily uniform), the overall
energy of a vortex should be unchanged if it is translated
by some amount. If this amount is small compared to the
coherence length (and this is certainly the case in
the experimental situation), then it can be described
by one quasiparticle-quasihole pair as given in section
\ref{sec:DISPLACED}. Thus one expects from translational
invariance that the residual interaction will reduce the
quasiparticle-quasihole excitation energy to nearly zero.

Unfortunately, an explicit demonstration of
final-state effects is difficult because in order to
get precisely zero energy one must possess precise
self-consistent order parameter and single-particle
states. This calculation would require detailed numerical
work even for a simple zero-range pairing interaction
and would be hopeless for any realistic model.
Nevertheless, let us show, analytically,
in a simple model, that
the correction to the excitation energy
is of order $-\Delta^{2}/(E_{F}{\rm ln}[2\omega_{c}/\Delta])$.
We begin with a simple zero-range pairing interaction
\begin{equation}
-V\int d{\bf r}
c_{\uparrow}^{\dagger}({\bf r})
c_{\uparrow}({\bf r})
c_{\downarrow}^{\dagger}({\bf r})
c_{\downarrow}({\bf r})\quad .
\end{equation}
In this case the residual interaction is
\begin{equation}
-V\int d{\bf r}
\left(
c_{\uparrow}^{\dagger}({\bf r})
c_{\downarrow}^{\dagger}({\bf r})
- \Delta^{*}({\bf r})
\right)
\left(
c_{\downarrow}({\bf r})
c_{\uparrow}({\bf r})
- \Delta({\bf r})
\right)\quad .
\end{equation}
This expression, converted to Bogoliubov operators, is
diagonal for the simplest $\mu = -1/2 \rightarrow +1/2$
excitation and has the expectation value
\begin{equation}
-V\int d{\bf r}
\left(
|u_{1/2}|^{2} + |v_{1/2}|^{2}
\right)^{2}\quad .
\end{equation}
Now using that fact that u and v have spatial extent
$\xi\sim\hbar v_{F}/\Delta$ and the simple BCS expression
$1=VN_{0}{\rm arcsinh}(\omega_{c}/\Delta)$ we arrive at the above
estimate. The main conclusion is that, in the vortex core,
it is possible for the residual interaction to
give negative corrections of the same
order of magnitude as the `bare' excitation energy.
In charged superconductors the collective mode
is `plasmonized' by the coulomb repulsion.
That is, there is a large positive correction to
the energy.
However this probably does not happen to the
quasiparticle-quasihole excitation in a vortex
because the corresponding translation of the vortex should not
create a charge density fluctuation.

In this paper we follow an alternative route and
explicitly allow the order parameter
to be dynamic, self-consistent and to affect the time evolution of
the quasiparticles. It is, in a sense, a conserving approximation
because it respects the translational invariance of the
problem.

To begin the calculation of conductivity we enumerate
the sources of electric field and current density and
consider only the uniform $q=0$ component.
The electric field due to the time derivative of the
background supercurrent is
${\bf\cal E} = (i\omega m/e){\bf v}_{S}$. There is
also the electric field due to transverse vortex motion,
$N_{v}(h/2e){\hat{\bf z}}{\bf\times}{\bf v}_{L}$,
where $N_{v}$ is the areal vortex density.
The trivial piece of the current density is simply the
supercurrent $n_{s}e{\bf v}_{S}$. Then there is the
$q=0$ component of the current density due to vortex motion,
from Eq. (\ref{PARAMAGNETIC}),
\begin{equation}
{\bf J}_{q=0} =
N_{v}
{
{e\hbar^{2}k_{F\perp}^{2}({\bf v}_{L}-{\bf v}_{S})}
\over
{2m(\Delta^{2}/E_{F})}
}\quad .
\end{equation}
We have added the term $-{\bf v}_{S}$ so that the total
current is $n_{s}ev_{S}$ when ${\bf v}_{L}={\bf v}_{S}$.
The next step is to write both the total electric field and
current density in terms of ${\bf v}_{S}$ using
the vortex equation of motion Eq.
(\ref{PARTICULAR}). Then the conductivity tensor, $\sigma_{ab}$,
may be found by comparing those two expressions and using the
definition $J_{a} = \sigma_{ab}{\cal E}_{b}$. The final result
for the conductivity tensor is rather complicated. To simplify it
we again take $n_{s}=k_{F\perp}^{2}/2\pi$,
$E_{F}=\hbar^{2}k_{F\perp}^{2}/2m$ and define $\Phi = N_{v}h/2m\Omega_{0}$
($\Phi$ is roughly $H/H_{c2}$).
The longitudinal and Hall conductivities are
\begin{equation}
\sigma_{xx}
=
{\cal F}
\left[
\left(
{{i\omega}\over{\Omega_{0}}}
+
{{\Omega_{0}\tau}\over D}
\Phi
\right)
\left(
1 - \Phi
{{1+i\omega\tau}\over D}
\right)
-
\Phi^{2}
{{\Omega_{0}\tau}\over D}
\left(
1 -
{{1+i\omega\tau}\over D}
\right)
\right]
\end{equation}
and
\begin{equation}
\sigma_{xy}
=
{\cal F}
\left[
\Phi
\left(
1 -
{{1+i\omega\tau}\over D}
\right)
\left(
1 - \Phi
{{1+i\omega\tau}\over D}
\right)
+
\Phi
{{\Omega_{0}\tau}\over D}
\left(
{{i\omega}\over{\Omega_{0}}}
+
{{\Omega_{0}\tau}\over D}
\Phi
\right)
\right]
\end{equation}
where $D\equiv (1+i\omega\tau)^{2} + \Omega_{0}^{2}\tau^{2}$
and
\begin{equation}
{\cal F} =
{{ne^{2}}\over{m\Omega_{0}}}
\left[
\left(
{{i\omega}\over{\Omega_{0}}}
+
{{\Omega_{0}\tau}\over D}
\Phi
\right)^{2}
+
\Phi^{2}
\left(
1 -
{{1+i\omega\tau}\over D}
\right)^{2}
\right]^{-1}.
\end{equation}

We are interested in the term in the conductivity
proportional to $\Phi$ or the number of vortices
when $\Phi$ is small.
However some care has to be taken with that limit because
in the low {\it frequency} limit the {\it resistivity} is
proportional to $\Phi$. Now, if we throw away terms of order $\Phi^{2}$
we have
\begin{equation}
\sigma_{xx}\sim
{{ne^{2}}\over{m\Omega_{0}}}
{
{
1 - \Phi
{{1+i\omega\tau}\over D}
}
\over
{
{{i\omega}\over{\Omega_{0}}}
+
{{\Omega_{0}\tau}\over D}
\Phi
}
}
\end{equation}
Note that in the $\omega\rightarrow 0$ limit this has the
correct value $\Phi^{-1}ne^{2}\tau/m$. In order to bring out the
structure, we take the $\tau\rightarrow\infty$ limit and obtain
\begin{equation}
\sigma_{xx}\sim
{{ne^{2}}\over{im\omega}}
{
\left[
	1 -
	{
	{i\omega^{2}\Phi}
	\over
	{\omega\tau(\Omega_{0}^{2}-\omega^{2})}
	}
\right]
\left[
	1 -
	{
	{i\Omega_{0}^{2}\Phi}
	\over
	{\omega\tau(\Omega_{0}^{2}-\omega^{2})}
	}
\right]
^{-1}\quad .
}
\end{equation}
At `resonance', $\omega\sim\Omega_{0}$, the
longitudinal conductivity becomes, in this limit,
completely imaginary. This agrees with the results of the previous subsection.

Similarly we may look analytically at ${\rm Im}\sigma_{xy}$ in the small
$\Phi$ limit. ${\rm Im}\sigma_{xy}$ gives the polarization dependent
dissipation. In this limit it has the value
\begin{equation}
{\rm Im}\sigma_{xy} \sim {{2ne^{2}\Phi\Omega_{0}\tau}\over{m\omega}}
\left[
\left(1 - \omega^{2}\tau^{2} + \Omega_{0}^{2}\tau^{2}\right)^{2}
+
(4\omega\tau)^{2}
\right]^{-1}.
\end{equation}
This expression has a pole-like feature
at $\omega\approx\Omega_{0}$ but it is weak
because its value at the maximum goes as
$1/\tau$ and therefore doesn't sharpen as
$\tau$ increases.

In Fig. (\ref{CONDUCTIVITY_PLOT}) we plot the  real
and imaginary parts of the longitudinal and Hall
conductivities as a function of frequency without
the above approximations.
They have been multiplied by $\omega$ or $\omega^{2}$
in order to bring out their behaviour near $\omega\sim\Omega_{0}$.
In general the conductivity at the characteristic frequency
does not show dramatic behaviour. What is important is that
${\rm Re}\sigma_{xx}$ does not show the expected peak
from a naive dipole resonance and ${\rm Im}\sigma_{xy}$ does
not show a strong peak as expected from a polarization
sensitive dipole resonance.
There is structure at low frequencies which depends on the value
of $\Phi$ but is not important to the
present discussion. It is related to the problem
with taking the low frequency limit. At some point
the {\it resistivity} becomes proportional to $\Phi$
instead of the conductivity. It is also the point at which the
electric field due to transverse vortex motion is comparable
to that of the time derivative of the gauge field.

\section{PINNING}
\label{sec:PINNING}

As mentioned in the introduction, in most practical cases pinning
comes into play. In this section we introduce the effects of
pinning phenomenologically and show that a dipole-like
resonance is recovered. First let us return to the homogeneous
solution of the equation of motion Eq. (\ref{HOMOGENEOUS}).
It should be possible to cause these modes to absorb energy,
however the equation of motion we derived apparently does not
allow it to happen resonantly.
It would seem that one requires a perturbation
to `bump' the system into one of these excited states.
That will be the purpose of the pinning center.

The main feature of pinning sites or
bending of vortices is that there is a
non-translation-invariant restoring force present.
Thus we propose to study the effect of modifying,
in a very general way, the
equation of motion, Eq. (\ref{MOTION}) into
\begin{equation}
{\dot{\bf v}}_{L}
=
{\dot{\bf v}}_{S}
+
\Omega_{0}
(
{\bf v}_{L}
-
{\bf v}_{S}
)
{\bf\times}
{\hat{\bf z}}
-
{1\over\tau}
{\bf v}_{L}
-
\alpha^{2}{\bf r}_{0}.
\end{equation}
Here ${\bf r}_{0}$ is the position of the vortex and
$\alpha$ is the characteristic frequency of a harmonic
well we have introduced at the origin.

Solving this equation
(and keeping only terms to order $1/\tau$)
we find the steady state solution to be
\begin{eqnarray}
\left(
\begin{array}{c}
v_{Lx}\\
v_{Ly}
\end{array}
\right)
=
i\omega
\left[
(\alpha^{2}-\omega^{2})^{2}
+
{{2i\omega}\over{\tau}}
(\alpha^{2}-\omega^{2})
-
\omega^{2}\Omega_{0}^{2}
\right]^{-1}\\
\times
\left(
\begin{array}{cc}
i\omega(\alpha^{2}+\Omega_{0}^{2}-\omega^{2} + i\omega/\tau)&
-\Omega_{0}(\alpha^{2} + i\omega/\tau)\\
\Omega_{0}(\alpha^{2} + i\omega/\tau)&
i\omega(\alpha^{2}+\Omega_{0}^{2}-\omega^{2} + i\omega/\tau)
\end{array}
\right)
\left(
\begin{array}{c}
v_{Sx}\\
v_{Sy}
\end{array}
\right)
\label{WITH_ALPHA}
\end{eqnarray}
Let us now also take $\alpha << \Omega_{0}$.
In that case the quantity in square brackets gives a resonance
at $\omega \approx \Omega_{0} + \alpha^{2}/\Omega_{0}$.
The term most important at resonance comes from the imaginary part
of the square bracket. Let us evaluate Eq. (\ref{WITH_ALPHA})
close to resonance by
setting $\omega = \Omega_{0}$ in other parts of the expression
to obtain
\begin{equation}
\left(
\begin{array}{c}
v_{Lx}\\
v_{Ly}
\end{array}
\right)
=
{
{\alpha^{2}\tau/(2\Omega_{0})}
\over
{1 + (\omega - \Omega_{0})^{2}\tau^{2}}
}
\left(
\begin{array}{cc}
-i & -1 \\
1 & -i
\end{array}
\right)
\left(
\begin{array}{c}
v_{Sx}\\
v_{Sy}
\end{array}
\right)\quad .
\end{equation}

The complete expression for the
conductivity is messy and perhaps not worth
writing down for this simple model.
However in the spirit of subsection \ref{subsec:SINGLE}
we may look at the dissipation in a single vortex.
Setting $v_{Sy} = v_{Sx}\exp{i\theta} = v_{s}/\sqrt{2}$
we may repeat the calculations leading to Eq. (\ref{SLIP})
and Eq. (\ref{DRUDE}) and obtain
\begin{equation}
{\rm Re}(v_{Lx}^{*}v_{Sy} - v_{Ly}^{*}v_{Sx}) = f (1 + {\rm
sin}\theta)v_{S}^{2}
\end{equation}
and
\begin{equation}
{\rm Re}(v_{Ly}^{*}i\omega v_{Sy} + v_{Lx}^{*}i\omega v_{Sx}) = -\omega f (1 +
{\rm sin}\theta)v_{S}^{2}
\end{equation}
where
\begin{equation}
f =
{
{\alpha^{2}\tau/(2\Omega_{0})}
\over
{1 + (\omega - \Omega_{0})^{2}\tau^{2}}
}
\end{equation}
gives the resonance structure which corresponds to the usual
dipole resonance.
Note that f is proportional to $\alpha^{2}$. The dipole absorption is
proportional to the strength of the translation-invariance-breaking
perturbation. This perturbation increases somewhat
the resonant frequency but need not dominate.
The resonance is also fully polarization dependent
and turns off when ${\rm sin}\theta = -1$, or, at the
correct choice of handedness of the circular
polarization.

\section{DISCUSSION}
\label{sec:DISCUSSION}

One of the main conclusions of this work, that in a pure system the
single-quasiparticle properties of a vortex are invisible to
a long wavelenth probe, is very reminiscent of the result of
Mattis and Bardeen \cite{MATTIS} regarding the long wavelength electromagnetic
response
near the bulk gap frequency. For vortices, if
$\tau\rightarrow\infty$ we have ${\bf v}_{L}={\bf v}_{S}$
and there is no dissipation. Mattis and Bardeen found that, in
the bulk, the
conductivity in the long wavelenth limit
$\sigma(q,\omega\sim 2\Delta)\rightarrow 0$ as
$q\rightarrow 0 $ unless there are impurities
present to allow the violation of momentum conservation.
There is a basic picture which is shared by these
two situations.
A quasi-particle-quasi-hole excitation can look like
a translation of the condensate as a whole and therefore
the system responds differently from what one expects.

In this paper we have calculated various quantities without the
need
to choose an appropriate `size' for the core.
In previous works \cite{BARDEEN_STEPHEN,NOZIERES} this
was required in order to fix the coefficient
of the `Lorentz' or `magnus' term. The
core size was chosen so that at $H=H_{c2}$ the core volume would be
the total volume of the system. Here there is no
particular reason to extrapolate to the superconducting
phase boundary.

The low temperature limit studied here may present
some simplifications.
A good definition
of the `core' of the vortex would be those
states closest to zero energy
which move at velocity ${\bf v}_{L}$. Some intermediate states
would have to be adjusted some other way to match the bulk
velocity ${\bf v}_{S}$ and conserve current.
They will be states further away from the
chemical potential and at low temperature perhaps
not important for calculating
dissipation. This point is discussed further in the
appendix. It is not clear at present exactly how crucial
it is.
The equation of motion of Bardeen-Stephen
is derived assuming that the full transport
current ${\bf v}_{S}$ flows through the vortex core and
this core is
necessarily defined by some surface or boundary region.
{}From the low temperature
microscopic point of view
the full transport current does not flow through the core.
That is because it is unfavourable to pay
the discrete amount of energy required to make charge
fluctuations in the core. Without these charge
fluctuations the paramagnetic current in the core is determined by
the velocity of the core because there aren't any other states
available in the vortex core.
At higher temperatures, it is
favourable to allow charge fluctuations in the
core and thus it is possible to increase the current flowing there.

Future work should concentrate on detailing the microscopic
theory of pinning and vortex bending since that would be most
important for the experiments. It would be very nice if the
de-pinning frequency correlation presented in table \ref{table}
could be explained. It would also be useful to work out
the quasiparticle structure of a vortex in a d-wave superconductor
That involves a non-local BdG equation. One could also check
how the discreteness of the lattice
affects the microscopic structure. In general, calculations of the
microscopic strcuture will be sensitive to the nature
of the pairing. Thus knowledge of the microscopic strcuture
of vortices
may give us some insight into the mechanism of high temperature
superconductivty.

\section{CONCLUSION}
\label{sec:CONCLUSION}

In this paper we have studied the microscopic dynamics of vortices
as a function of frequency. We have derived an equation
of motion and calculated the electrodynamical
response at $q=0$. For free vortices we find that
the single-particle character is essentially
invisible unless translation invariance is broken.
Thus the
clear, albeit broad, response in optical
transmission of Karra\"i {\it et al.} is
perhaps due to pinned vortices.
Another
interesting possibility is that
interaction with the discrete lattice is responsible.
The possibly d-wave
nature of the paired state in high temperature superconductors
should not affect this conclusion. We believe that the
large width observed experimentally may come from intrinsic
scattering from whatever breaks translation invariance.

\acknowledgements

The author wishes to acknowledge conversations with
H.D. Drew, W. Hardy, K. Karra\"i, J. Rammer, P. Schleger and N. Schopol.

\unletteredappendix{Matrix elements}

Throughout this paper we use matrix elements of the quasiparticle
wavefunctions. They are difficult to calculate because the
detailed behaviour of the wavefunctions depends non-trivially on the
radial dependence of $|\Delta({\bf r})|$ which in turn
is determined self-consistently.
In this appendix we derive approximate matrix
elements of ${\bf r}$ and ${\bf\nabla}$ between the low energy
quasiparticle states. The approximation we make is that
the main contribution to these matrix elements comes from
close to the center of the vortex where the functions $u({\bf r})$ and
$v({\bf r})$ can be approximated by Bessel functions and the
order parameter $\Delta({\bf r})$ is small.
It is assumed that matrix elements of the low energy states
are not very sensitive to
the exact behaviour of the wavefunctions near the core
boundary because the functions u and v are
suppressed exponentially when the magnitude $|\Delta({\bf r})|$
becomes substantial. Boundary conditions have entered implicitly
through the energies of the quasiparticle levels.

We shall make use of the Bessel function identities
\begin{eqnarray}
J_{\nu - 1}(z) + J_{\nu + 1}(z) &= {{2\nu}\over z}J_{\nu}(z),\nonumber\\
J_{\nu - 1}(z) - J_{\nu + 1}(z) &= 2{d\over{dz}}J_{\nu}(z).
\end{eqnarray}
and suppose that $u({\bf r})$ and $v({\bf r})$ satisfy similar
identities at least where they contribute most to the matrix element.
To be specific, in the calculation of matrix elements of $\bf\nabla$
we use
\begin{eqnarray}
{
{\partial}
\over
{\partial r}
}
u_{\mu}
=&
{{k_{F\perp}} \over {2}}
\left(
e^{i\theta}u_{\mu -1} - e^{-i\theta}u_{\mu + 1}
\right)\nonumber\\
{1\over r}
{
{\partial}
\over
{\partial \theta}
}
u_{\mu}
=&
-{{k_{F\perp}}\over 2i}
\left(
e^{i\theta}u_{\mu -1} + e^{-i\theta}u_{\mu + 1}
\right)
\label{GRAD_UV}
\end{eqnarray}
and exactly the same equations with u replaced by v.
$\theta$ is the polar angle.
The matrix elements of
\begin{equation}
{\hat{\bf n}}\cdot{\bf\nabla} = {\rm cos}(\theta-\theta_{0})
{{\partial}\over{\partial r}} - {\rm sin}(\theta-\theta_{0})
{{\partial}\over{r\partial\theta}}
\end{equation}
where ${\hat{\bf n}}$ is a unit vector pointing at angle $\theta_{0}$ to
the x-axis, are thus
\begin{equation}
\int d^{3}{\bf r} \psi_{\mu\pm 1}^{\dagger}
{\hat{\bf n}}\cdot{\bf\nabla}
\psi_{\mu}
=
{{k_{F\perp}}\over{2}} e^{\mp i(\theta_{0}+(\pi/2))}
\end{equation}
and zero for states differing by other angular momenta.
The lack of dependence on $\mu$ is
a result of the approximations we have made.

Another approximation we have made is to only consider
matrix elements between low energy states. We consider
in this paper mostly operators that contain angular momentum
$l=\pm 1$ and therefore amongst the low energy states it is
clear which matrix elements are non-zero. The approximation
made in Eq. (\ref{GRAD_UV}) does not affect that. There are,
however, high energy scattering states (of energy $\Delta$ or higher)
which have non-zero matrix elements. These have been ignored.
It is not clear whether this is a safe approximation or not.

Now let us consider the matrix elements of
$\sigma^{z}{\hat{\bf n}}\cdot{\bf r}$.
To do this we consider the commutator
$[H,\sigma^{z}{\hat{\bf n}}\cdot{\bf r}]$
where H is the BdG Hamiltonian with ${\bf A}=0$.
We have
\begin{eqnarray}
\int \psi_{\mu}^{\dagger} \sigma^{z}{\hat{\bf n}}\cdot{\bf r} \psi_{\nu}
=& {1\over{\epsilon_{\mu}-\epsilon_{\nu}}}
\int &\psi_{\mu}^{\dagger}
[H,\sigma^{z}{\hat{\bf n}}\cdot{\bf r}] \psi_{\nu}
\nonumber\\
=& {1\over{\epsilon_{\mu}-\epsilon_{\nu}}}
\int &\psi_{\mu}^{\dagger}
\left[
{-i\over m}{\hat{\bf n}}\cdot{\bf p}
+ 2{\hat{\bf n}}
\left(
\begin{array}{lr}
0 & -\Delta \\
\Delta^{*} & 0
\end{array}
\right)
\right]
\psi_{\nu}
\end{eqnarray}

The second term in the square brackets is a smaller contribution
because whenever u and v are substantial, $|\Delta|$ is small
so in our approximation we shall ignore it.
We are then left with the matrix element of the momentum operator
which we have just evaluated above.
The matrix element is
\begin{equation}
{{-\hbar k_{F\perp}}\over {2m(\epsilon_{\mu}-\epsilon_{\nu})}}
\left(
e^{i\theta_{0}}\delta_{\mu,\nu -1}
-
e^{-i\theta_{0}}\delta_{\mu,\nu +1}
\right)\quad .
\end{equation}

In section \ref{sec:RESPONSE}
we encounter the matrix element
\begin{equation}
W_{\mu\nu} = \int d^{3}{\bf r} ~\psi_{\mu}^{\dagger}({\bf r})
\left(
\begin{array}{cc}
0 &\delta\Delta\\
\delta\Delta^{*}&0
\end{array}
\right)
\psi_{\nu}({\bf r})\quad .
\end{equation}
where
$\delta\Delta = -\delta{\bf r}_{0}^{\prime}\cdot{\bf\nabla}\Delta$.
It may be found by utilizing translation invariance
if $\mu\ne\nu$.
Let
$\delta\psi_{\nu}({\bf r}) = -\delta{\bf
r}_{0}^{\prime}\cdot{\bf\nabla}\psi_{\nu}({\bf r})$
be the change in the
eigenfunctions upon displacing them by
an amount $\delta {\bf r}_{0}^{\prime}$.
Then substituting $\delta\psi_{\nu}$ and $\delta\Delta$ into
the Schr\"odinger equation, Eq. (\ref{BGD}),
subtracting the same equation with
$\delta{\bf r}_{0}^{\prime}=0$,
keeping terms to first order in
$\delta {\bf r}_{0}^{\prime}$, multiplying
on the left by $\psi_{\mu}({\bf r})$ and
integrating by parts results in
\begin{equation}
W_{\mu\nu} = (\epsilon_{\nu} - \epsilon_{\mu})
\int d^{3}{\bf r} ~\psi_{\mu}^{\dagger}({\bf r})
\delta\psi_{\nu}({\bf r})\quad .
\end{equation}
This can be further simplified making the assumption
again that $u({\bf r})$ and $v({\bf r})$ resemble
Bessel functions in the relevant spatial region
yielding
\begin{equation}
\delta {\bf r}_{0}^{\prime}\cdot{\bf\nabla}
\psi_{\mu}({\bf r}_{0}^{\prime})
=
{{\delta rk_{F\perp}}\over{2i}}
\left[
e^{i(\phi_{0}^{\prime} + {\pi\over 2})}\psi_{\mu - 1}({\bf r})
+
e^{-i(\phi_{0}^{\prime} + {\pi\over 2})}\psi_{\mu + 1}({\bf r})
\right]
\end{equation}
where $\delta{\bf r}$ makes an angle $\phi$
with the ${\hat x}$ axis.
In terms of the displacement $\delta {\bf r}_{0}^{\prime}$
the matrix element is,
\begin{equation}
W_{\mu\nu} = -{
{\delta r_{0}^{\prime}k_{F\perp}}
\over
{2i}
}
\delta_{\mu ,\nu\mp 1}
(\epsilon_{\nu} - \epsilon_{\mu})
e^{\pm i(\phi_{0}^{\prime} + {\pi\over 2})}\quad .
\end{equation}

%
%
\figure{(a) Dissipation (normalized to the high frequency limit)
	as a function of frequency (normalized to $\Omega_{0}$)
	when $\Omega_{0}\tau = 1$. Solid line: circular polarization
	of ${\bf v}_{S}$; dashed line: opposite sense of
	circular polarization. (b) The same quantities but for
	$\Omega_{0}\tau = 3$.
\label{DISSIPATION_PLOT}}
\figure{(a) Real part of the longitudinal conductivity multiplied
	by frequency $\omega$ (for clarity), ${\rm Re}\omega\sigma^{xx}$,
	as a function of frequency (normalized to $\Omega_{0}$).
	Solid line: $\Phi=0.1$; dashed line: $\Phi=0.2$ . (b) Same
	plots for the imaginary part multiplied by
	$\omega^{2}$ for clarity, ${\rm Im}\omega^{2}\sigma^{xx}$.
	(c) Real part of the Hall conductivity,
	${\rm Re}\omega^{2}\sigma^{xy}$.
	(d) Imaginary part, ${\rm Im}\omega^{2}\sigma^{xy}$.
\label{CONDUCTIVITY_PLOT}}
%
%
\begin{table}
\caption{Observed crossover frequencies for vortex de-pinning and
quasiparticle energy level separations in vortex cores ($\Delta^{2}/E_{F}$,
where $\Delta$ is the bulk gap and $E_{F}$ is the Fermi energy) for
various samples of low temperature superconductors.}\label{table}
\begin{tabular}{lccc}
\ & $T_{c} (K)$ & Observed frequency (MHz) &
	Energy level separation (MHz)\\
\tableline
$\rm{PbIn}$ \cite{GITTLEMANA} & 1.7 & 4 & 2 \\
$\rm{PbIn}$ \cite{GITTLEMANA} & 1.7 & 5 & 2 \\
$\rm{NbTa}$ \cite{GITTLEMANA} & 4.2 & 15 & 20 \\
$\rm{PbIn}$ \cite{GITTLEMANB} & 1.7 & 7 & 2 \\
$\rm{PbIn}$ \cite{GITTLEMANB} & 1.7 & 8 & 2 \\
$\rm{NbTa}$ \cite{GITTLEMANB} & 4.2 & 26 & 20 \\
$\rm{NbSe_{2}}$ \cite{SCHLEGER} & 7.2 & 100 & 1000 \\
\end{tabular}
\end{table}

\begin{references}
\bibitem{MISTAKE} The polarization dependent term of this
	component was omitted in the earlier paper Ref.
	\cite{HSU}.
\end{references}
\end{document}